\documentclass{emulateapj}

\slugcomment{To appear in ApJ Letters}

\shorttitle{Galactic Center}
\shortauthors{Laycock et al.}

\begin{document}


\title{Constraining the Nature of the Galactic Center X-ray Source Population}


\author{S. Laycock\altaffilmark{1}, J. Grindlay\altaffilmark{1}, M. van den Berg\altaffilmark{1}, 
P. Zhao\altaffilmark{1}, J. Hong\altaffilmark{1}, X. Koenig\altaffilmark{1}, 
E. M. Schlegel\altaffilmark{1}, S. E. Persson\altaffilmark{2}}

\altaffiltext{1}{Harvard-Smithsonian Center for Astrophysics, 60 Garden St, Cambridge, MA, 02138} 
\altaffiltext{2}{Observatories of the Carnegie Institution of Washington, 813 Santa Barbara St.,
Pasadena, CA, 91101}



\begin{abstract}
We searched for infrared counterparts to the cluster of X-ray
point sources discovered by Chandra in the Galactic Center Region (GCR).
While the sources could be white dwarfs, neutron stars, or black holes
accreting from stellar companions, their X-ray properties are
consistent with magnetic Cataclysmic Variables, or High Mass
X-ray Binaries (HMXB) at low accretion-rates. A direct way to decide 
between these possibilities and hence between alternative formation scenarios
is to measure or constrain the luminosity distribution of the
companions. Using infrared (J, H, K, Br${\gamma}$) imaging, we
searched for counterparts corresponding to typical HMXB secondaries:
spectral type B0V with K$<$15 at the GCR. We found no significant
excess of bright stars in Chandra error circles, indicating that HMXBs
are not the dominant X-ray source population, and account for fewer than 
10\% of the hardest X-ray sources.
\end{abstract}




\section{Introduction}
With the high spatial resolution of Chandra and ground-based infrared
(IR) instrumentation, the Galactic Center, heavily obscured by dust,
is now accessible to a wide range of astrophysical investigations.
A series of Chandra observations of the GCR ($\sim$10$'$ region around SgrA*), 
totaling 670 ksec, have recently been analyzed by \cite{muno2003}(hereafter, M03), revealing
2357 point sources. The 2-8 keV X-ray luminosity ($10^{31-33}$ erg s$^{-1}$)
and spectral index ($\Gamma < 1$) of the majority of the sources
are inconsistent with normal stars, active binaries, or
young stellar objects, none of which are normally persistently luminous enough 
at $>$2.5 keV to produce the
observed fluxes in the presence of the high absorption 
column (N$_{H}\sim 10^{23} cm^{-2}$). The only known sources expected to be detected
given the absorption are compact objects accreting from binary
companions. The luminosity range of the GCR sources covers magnetic CVs
and non-Roche-lobe filling HMXBs accreting in a low mass-transfer
regime, e.g.~X~Per, A0535+26 \citep{negueruela2000}. 
Quiescent low mass X-ray binaries (LMXBs) with neutron star (NS) or black hole (BH) 
primaries are possible contributors but are generally too faint (BHs) or too soft (NSs).
HMXBs are young objects (few My), while CVs are slow to form and long-lived.  The two
possibilities point to different evolutionary phases of the GCR.

If HMXBs comprise a significant fraction of the GCR sources, they
must be the products of recent, massive-star
formation. Frequent star-burst events in the GCR are expected on
$\sim$20 My scales based upon the instability of large gas clouds
orbiting within 200pc of the nucleus \citep{stark2004} and the
presence of several young clusters of massive stars. The HMXB hypothesis 
 was proposed and discussed in detail by \cite{pfahl2002}.

\cite{grindlay1985} suggested that the Galactic Bulge contains the
debris of tidally-disrupted globular clusters.  Were the GCR sources
dominated by magnetic CVs, globular clusters would then contribute
because CVs are over-produced per unit mass in globulars relative 
to the Galactic plane.

\cite{morris1993} proposed a third scenario in which compact stellar
remnants form in the Galactic bulge and sink toward the center of the
dense mass distribution via gravitational interactions.  The presence
of a super-massive black hole (BH) at the center accelerates the
process, leading to a central density cusp. This scenario fits with
the \cite{muno2005} (M05) discovery of a centrally concentrated
overabundance of X-ray transients, as well as the overall 1/$\theta$ 
source distribution found by M03. 

All three channels may be active now or have been in the past, so by measuring
relative numbers of the various end-products, we hope to piece
together the composition and history of the GCR.  In this Letter we
present preliminary results of our deep IR survey targeted at
measuring the HMXB fraction. The occurrence of IR counterparts
consistent with highly reddened massive stars and giants will constrain the
relative abundance of HMXBs and CVs. This work is part of our Chandra 
Multi-wavelength Plane Survey (ChaMPlane: \citealt{grindlay2005}) 
which is an effort to map the space density of all types of low-luminosity 
accretion sources in the Galaxy.

\section{Observations}
The GCR is heavily obscured with average A$_{V}$$\sim$25 so counterparts can
only be detected with IR imaging.  The stellar density is also very
high so that reaching a sufficient limiting magnitude to make meaningful
constraints on population models demands excellent image quality. In
June 2004 we observed the inner 10$'$x10$'$ of the GCR with the PANIC
imager on the 6.5m Magellan (Baade) telescope at Las Campanas,
Chile. PANIC has a 2$'$x2$'$ field of view with 0.125$''$
pixels. Under good seeing conditions (median FWHM $\sim$ 0.5$''$), we
obtained a mosaic of 25 pointings in J, H, Ks, and Br$\gamma$ filters
using a 5-point dither pattern with 3 exposures at each position. 
Sky emission was removed using off-source exposures, and the 
images were reduced and stacked in IRAF.
Photometry was performed on the stacked
images using SExtractor \citep{bertin1996}. Astrometric and
photometric calibration were performed using the 2MASS point-source
catalog. 

We used a version of X-ray point source catalog of M03, kindly provided by M. Muno,
listing J2000 co-ordinates with net count rate, flux, and source
significance determined in 3 energy bands S1 (0.3-1.2keV), S2 (1.2-2.5
keV), and H (2.5-8 keV). The intermediate S2 band provides some
additional information, but for this Letter we restrict ourselves to
discussion of GCR (H only) and foreground (S1) sources.

\section{Experimental Design}
\label{sect:design}
We set out to identify all potential IR counterparts to the
M03 X-ray sources and apply statistical tests to measure their numbers relative to 
random positional coincidence. Our scientific goals are to 
set stringent upper-limits on the number of counterparts and examine their photometric properties.

The availability of X-ray spectral information and multi-band IR
photometry permits a clean separation of sources situated in the GCR
from those in the foreground. The GCR sources are characterized by
X-ray spectra completely cut off at low energies due to absorbing gas
and dust. Of the sources in the M03 catalog, 80\% are detected at
$>$3$\sigma$ in the H band only, while any source detected in the
S1 band must lie in the foreground. The spatial
distributions of X-ray point sources and K-band stars follow similar
radial profiles about the GCR. M03 demonstrated a surface density with
1/$\theta$ dependence with angular offset $\theta$ from SgrA*.
Similar results have been reported for stars from IR 
observations (e.g. \citealt{catchpole1990}).

There are $\sim$100 known Galactic HMXBs, of which 70\% have Be star
companions whose spectral types lie in the narrow range O9-B2 and peak
at B0V \citep{negueruela1998}.  We adopt the luminosity and color of a
B0V to represent ``typical'' HMXB values ($M_{V}$=$-$4, V$-$K=$-$0.83;
\citealt{allen}).  Note that many non-Be HMXBs are {\it more}
luminous because they have super-giant companions. For the CV case, we
adopt a K0V dwarf (M$_{V}$=5.9,V$-$K=1.96) at the brighter
end of the normal range \citep{patterson1998}. Adopting $A_{V}$=25 
and $D$=8 kpc for the GCR, our target HMXB should have an
apparent magnitude K=14.4 while the assumed CV would have K=21. We note the 
intermediate polar GK Per with bright sub-giant counterpart would have K$\sim$19.
 
\subsection{Photometric Completeness Curve}
\begin{figure}
\includegraphics[angle=-90,width=7.5cm]{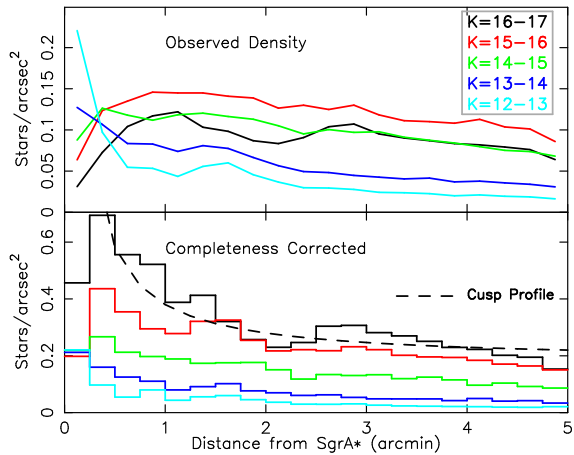}
\caption{Star density curves for PANIC K-band photometry, 
in 5 magnitude ranges as indicated by legend.
Observed number-density is shown by smooth lines, completeness corrected shown by histogram.
A radial profile of the form A+B$\times\theta^{-1}$ is also shown.
\label{fig:photcom}
}
\end{figure}

Due to severe crowding in the GCR, we infer the star density from our observations
by correcting for incompleteness i.e. the loss of faint stars due to their brighter neighbors. 
Completeness curves for our PANIC photometry as functions of magnitude 
and distance from SgrA* were determined by simulation.
For each of the 25 K-band stacked images, a random sample of stars was added
using IRAF {\it mkstar} with a Gaussian PSF matched to the seeing. The 
photometry was repeated and the artificial stars identified and compared to 
their input magnitudes. A master catalog of input and recovered stars was 
then assembled, from which completeness curves could be extracted for any 
region of the 10$'$x10$'$ mosaic. The 50\% completeness limit (fraction of stars 
recovered within $\pm$0.5 mag of their input magnitude) within the inner 1$'$ 
radius occurs at K=14.5 and over the rest of the mosaic at K=15.4. 
Curves for star density are presented in Figure~\ref{fig:photcom} and 
completeness values for the full field reported in the final column of Table~\ref{tab:peakupresults}. 

\subsection{Peak-up Test} 
The IR and X-ray catalogs are repeatedly matched
over a large grid of positional offsets using integer multiples of the
PANIC pixel-size of 0.125$''$. At each offset position, the number of
X-ray sources with one or more stars lying in their error-circle is
recorded in a series of magnitude bins.
We used 95\% confidence error-circles determined from 
the formula of \cite{hong2005} which is dependent on net counts and 
distance from the Chandra aimpoint. For the GCR sources, error radii 
are 0.3$''$-1$''$. The presence of real counterparts is signaled
by a peak at the offset co-ordinates corresponding to perfect
alignment of the two catalogs, providing a boresight. 
To confirm that X-ray and IR catalogs are on the same astrometric
frame we centroided the soft-source peak, finding a residual offset 
of ${\Delta}$RA=0.011(1)$''$, ${\Delta}$Dec=0.092(1)$''$ with FWHM=0.55$''$.

\begin{figure}
\includegraphics[angle=-90,width=7.5cm]{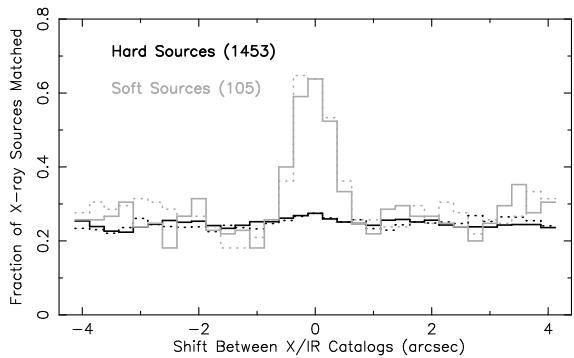}
\caption{Peak-Up test for hard and soft X-ray sources, using PANIC K-band photometry. 
Solid and dotted lines show projections along the RA and Dec axes of the number of X-ray 
sources with one or more stars inside the 95\% error circle. A prominent peak is seen at
zero offset only for the soft sources. The random match level and its fluctuations 
are apparent beyond 1$''$ offset.
\label{fig:peakup}
}
\end{figure}

The number of random matches ($N_{ran}$) is calculated as the 
mean of all trials at shifts greater than 2$''$, i.e. far away from correct alignment, 
and the standard deviation ($\sigma$) of these values is used to characterize the
uncertainty. We then take the (peak) number of matches ($N_{m}$), and subtract the random level to 
obtain the excess-above-random ($N_{exc}$). We claim detection of a population of 
counterparts above that predicted to occur by chance if $N_{exc}$ exceeds $3 \times \sigma$.
The upper-limit on the total number of true counterparts was determined by monte-carlo simulation.
At each iteration the X-ray catalog was offset by randomly generated shifts in X and Y such that 
the absolute shift was between 2$''$ and 5$''$. A randomly selected sample of $N_{true}$ X-ray sources were 
flagged as ``true counterparts'' and the catalogs matched as previously described. Since all of 
the X/IR matches were by definition random (due to the offsets) we were able to count the total 
number of matches $N_{sim}$ in each magnitude bin, correcting for duplicates (sources flagged as 
simulated matches which also happened to have a random IR counterpart) such that 
$N_{sim}$ = $N_{ran}$ + $N_{true}$ - $N_{ran.AND.true}$.
Guided by the values of $\sigma$ from the peakup test, we ran simulations with $N_{true}$=5-40 
with 100 iterations at each value. For each value of $N_{true}$ we recorded the largest 
value of $N_{sim}$ reached in 90\% of trials, and fitted the results with a straight line relationship.
Thus the upper-limit on the number of true matches corresponds to the value of $N_{true}$ when
$N_{sim}$ = $N_{m}$, these values are given in Table~\ref{tab:peakupresults} as $U_{90}$.

\section{Constraints on IR/X-ray Counterparts}
We performed the peak-up test first on the entire
field, and then independently on a series of concentric annular regions centered on SgrA*. 
This allows us to limit the effects of crowding in the IR, and to follow 
the cusp distribution of hard X-ray sources. 

For the full field we found no significant excess in the number of IR matches for hard X-ray 
sources, irrespective of magnitude, while for soft sources we found a highly significant 
($\sim$30$\sigma$) counterpart population dominated by bright (K$\sim$11) stars. 
This result is illustrated visually by Figure~\ref{fig:peakup} and in 
more detail in Table~\ref{tab:peakupresults}. There is a marginal signal ($2.9\sigma$) for 
hard sources versus the entire PANIC catalog, which could not be associated with any specific 
magnitude range. 

\begin{deluxetable}{lrrrrrrrr}
\tabletypesize{\scriptsize}
\tablecaption{Peakup Test on the full 10$'$x10$'$ field. \label{tab:peakupresults}}
\tablehead{
\colhead{K$_1$-K$_2$} & \colhead{$N_m$}   & \colhead{$N_{ran}$} & \colhead{$\sigma$} &
\colhead{$N_{exc}$}  & \colhead{Sig} & \colhead{C} & \colhead{{\it U$_{90}$}} & \colhead{{\it U$_{c}$}}
}
\startdata
\multicolumn{3}{l}{Hard Sources (1453)} \\
$>$12      & 399 & 353  & 15.8 & 46  & 2.9  & -    & 87$^{*}$  & -   \\
$>$17      & 86  & 77.9 & 8.1  & 8.1 & 1.0  & -    & 20$^{*}$  &  -  \\
16-17      & 72  & 57.7 & 7.2  & 14.3& 2.0  & 0.3  & 54  & 10.7 \\
15-16      & 84  & 83.0 & 9.3  & 1.0 & 0.1  & 0.5  & 27  & 7.3 \\
14-15      & 83  & 67.5 & 8.3  & 15.5& 1.9  & 0.7  & 37  & 5.4 \\
13-14      & 37  & 35.7 & 6.1  & 1.3 & 0.2  & 0.8  & 10  & 3.0 \\
12-13      & 30  & 22.6 & 4.5  & 7.4 & 1.6  & 0.9  & 13  & 2.3 \\
11-12      & 11  & 8.6  & 2.4  & 2.4 & 1.0  & 1.0  & 11  & 1.4 \\
10-11      & 6   & 5.3  & 1.9  & 0.7 & 0.4  & 1.0  & 6   & 0.6 \\
9-10       & 2   & 2.2  & 1.3  & -0.2& -0.2 & 1.0  & 2   & 0.2 \\
\multicolumn{3}{l}{Soft Sources (105)}  \\
$>$12      & 67  & 28.6 & 4.7  & 38.4& 8.2  & -    & 59$^{*}$   & -  \\  
15-16      & 8   & 6.2  & 2.6  & 1.8 & 0.7  & 0.5  & 10  & 92  \\
14-15      & 21  & 5.3  & 2.3  & 15.7& 6.8  & 0.7  & 27  & 84  \\
13-14      & 10  & 2.6  & 1.4  & 7.4 & 5.3  & 0.8  & 11  & 59  \\
12-13      & 13  & 1.5  & 1.3  & 11.5& 8.8  & 0.9  & 14  & 49  \\
11-12      & 21  & 0.4  & 0.7  & 20.6& 29.4 & 1.0  & 21  & 35  \\
10-11      & 12  & 0.2  & 0.5  & 11.8& 23.6 & 1.0  & 12  & 15  \\
9-10       & 4   & 0.0  & 0.1  & 4.0 & 40   & 1.0  & 4   & 4 
\enddata
\tablecomments{{\tt In the $\sim$ 10$'$x10$'$ PANIC survey area we consider: 184$\times10^3$
K-band stars, 1453 hard X-ray sources, 105 soft X-ray sources.
2MASS point-source catalog used for K$<$12.
{\em $N_m$}= Number of X-ray sources with one or more matches.
{\em $N_{ran}$}= random matches, {\em $\sigma$} uncertainty in number of random matches.
{\em $N_{exc}$}= number of "excess" matches above random level.
{\em Sig} significance of excess.
{\em C} Completeness fraction for K-band stars.
{\em {\it U$_{90}$}}= Completeness corrected upper-limit on the number of counterparts at 90\% significance.
{\em {\it U$_{c}$}}= Cumulative {\it U$_{90}$} brightward of K$_2$ expressed as \% of number of X-ray sources.
*No completeness correction applied for unbounded mag ranges.
}}
\end{deluxetable}

We therefore conclude the GCR counterpart population is dominated by stars fainter
than our confusion limit, although some fraction of the hard matches are undoubtedly real. 
The 90\% confidence upper limit for the full catalog indicates there are 
up to 87 hard counterparts although we cannot at this stage identify them 
from among the 399 total matches.
By placing completeness-corrected upper-limits on the number of counterparts as a function of K magnitude, 
we can constrain the proportion of HMXBs, and hence place a lower-limit on the population of CVs and LMXBs.  
The final column of Table~\ref{tab:peakupresults} gives the cumulative percentage obtained by adding 
the upper-limits for each magnitude range and dividing by the number of X-ray sources. We find 
that a maximum of 5.4\% (at 90\% confidence) of the sources could have counterparts brighter than K=15, 
given average GCR extinction. The presence of a significant HMXB population would appear between K=14-16, and the 
fact that we {\it do} detect soft-source counterparts for K=14-15 (at 6.8$\sigma$) demonstrates 
that we are not merely being defeated by crowding in this range.

For annular bins 0-1$'$, 1$'$-2$'$, 2$'$-3$'$, 3$'$-4$'$, 4$'$-5$'$, our results constrain the 
HMXB population as a function of distance from SgrA*. The effect of the nuclear cluster is 
apparent in the inner 1$'$, where a large number of massive stars raise the upper-limits such 
that K$<$15 HMXB can potentially account for 9\% of the 110 hard X-ray sources present 
(at 90\% confidence). The limit is 5-8\% for the remaining annuli. Overall the annular results 
are less constraining than the full field results due to the smaller sample sizes involved. 
A single annular bin (2$'$-3$'$) produced a peak 
at 3.8$\sigma$ significance, possibly corresponding to a drop in the confusion 
limit sufficient to allow counterparts to be isolated. Such an effect could be 
due to seeing, anomalous extinction or an unrelated star cluster. 

We can account for 84\% of the 105 soft sources down to K$<$15, and essentially all to K$<$16 where 
crowding dominates. The majority of soft sources will be dMe and other foreground coronal stars. 
For example at spectral type M0V our 50\% completeness limit of K$\sim$15.4 implies a distance of just 1kpc, 
indeed 86\% of the S1 sources have counterparts from our optical ChaMPlane survey \cite{zhao2003}.

\section{Photometry of Potential Counterparts} 
\begin{figure}
\includegraphics[angle=-90,width=8.5cm]{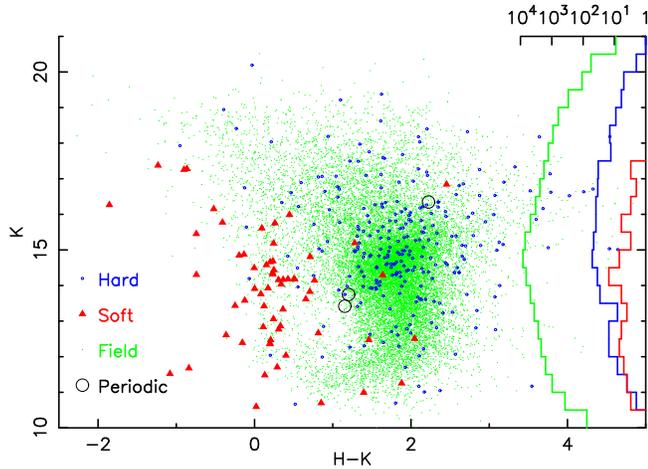}
\caption{Color-Magnitude diagram comparing the IR counterparts of hard
and soft X-ray sources with the field population. Total numbers in each 
class given in histogram at right. (Saturation begins at K$\lesssim$11.5)
\label{fig:colormag}}
\end{figure}

Color-magnitude diagrams (CMD) were constructed to compare the H--K vs K 
distribution for stars lying in X-ray error-circles with that of 
the field population. Extinction due to dust lying between us and the 
Galactic Center makes the GCR stars appear systematically redder than 
foreground stars, largely overcoming any degeneracy between intrinsic 
brightness and distance.
In figure~\ref{fig:colormag} we see the soft counterparts are mainly
un-reddened stars, while the hard counterparts cluster around H--K$\sim$1.6 with a large scatter. 
Intrinsic H--K colors do not exceed $\sim$+0.3 for any stars (\citealt{allen}).
Soft-source counterparts are also brighter than average field stars,
the K-S probability that they are the same is 2.6$\times 10^{-9}$. 
This result is consistent with the X-ray selection and peak-up test, the 
soft sources are shown to lie in the foreground, while the hard sources 
are dominated by random matches to stars in the GCR.  

Matches were found for 3 out of the 8 periodic X-ray sources found by \cite{muno2004} (7 in PANIC field, 5
observed in Ks, see Fig.~\ref{fig:colormag}). One of these (CXOGCR J174534.5-290201) is
heavily reddened, while the others have slightly lower H--K values
than the main concentration of stars in the CMD. 
The colors of the candidate stars do not agree with the predicted extinction from X-ray
spectral fits of \cite{muno2004}. The X-ray derived N$_{H}$ of
$\gtrsim$10$^{23}$cm$^{-2}$ requires the stars to be extremely bright to
explain the observed K magnitudes and predicts H--K colors 2 mag
redder than what is observed. These observations may be reconciled if most
of the X-ray-derived N$_{H}$ arises in the binary, in an accretion column or disk obscuring 
the X-ray source. A column of a few 10$^{22}$cm$^{-2}$ is typical for HMXB pulsars, but is
insufficient to make up the difference. It is also possible that these matches are random 
and the real counterparts are much fainter. No match was found for the 7 bright transients detected by M05.

A preliminary search for Br$\gamma$ emission objects among
X-ray IR counterparts yielded 11 candidates with S/N$>$5, of which 3
are foreground objects. Br$\gamma$--K is severely restricted by the
small amplitude of the expected signal ($\sim$-0.1 mag for EW(Br$\gamma$)$\sim$10-20\AA) and the
effects of crowding.

\section{Conclusions} 
The highly absorbed GCR X-ray sources appear to be
dominated by a population of accreting binaries in which the
mass-donor is less massive than in a typical HMXB. This conclusion follows
from the lack of a significant excess of stars brighter than K$\lesssim$15 in the
error circles of hard X-ray sources, which is the apparent magnitude of a B0V
star at the distance and mean Av of the GCR. By applying our peak-up test as a function of magnitude and 
distance from SgrA*, and correcting for completeness, we set 90\% confidence upper limits on the number of 
hard X-ray sources that can have counterparts in a series of ranges between K=9-16. After accumulating 
these limits for all counterparts brighter than K=15, such stars can account for at most 
79 out of 1453 hard X-ray sources, or 5.4\%. Pushing one magnitude fainter to include Be 
HMXBs as late as B2, we find an upper-limit of 103 or 7.1\%. Within 1$'$ of SgrA* these upper-limits become 
9\% and 15\% due to the high density of very luminous stars in the nuclear cluster. We note that large 
variations in A$_v$ and/or star clustering could modify these numbers.

Our constraints imply non-HMXBs account for more than 90\% of the M03 X-ray sources (at least beyond 1$'$)
favoring the leading alternative hypothesis of magnetic CVs. High angular resolution spectroscopy will be 
essential to identify the true counterparts which likely exhibit B$\gamma$ or
He emission lines indicative of an accretion disk.

This work was funded in part by Chandra grant AR4-5003A and NSF grant AST-0098683.
We thank the referee M. Muno for his insightful suggestions.

\end{document}